TEX, C Version 3.14t3
\documentstyle[12pt]{article}
\title{\bf
             CURIE TEMPERATURES \\ FOR BINARY
             ISING FERROMAGNETS \\ ON THE SQUARE LATTICE}
\author{
{\bf   Z. N\'eda } \\
{\small\it
      Babes-B\'olyai University, Dept. of Physics}   \\
{\small\it
      str. Kogalniceanu 1, RO-3400 Cluj, Romania  }   \\
     \and
{\small\it and  }\\
{\small\it
       University of Bergen, Dept. of Physics   }\\
{\small\it
         All\'egaten 55, N-5007 Bergen, Norway}  }
\date{}

\begin{document}

\maketitle

\begin{center}

Abstract\\
\end{center}

High-accuracy Swendsen and Wang Monte Carlo simulations were performed to study
the Curie temperature of ferromagnetic, binary Ising systems on the square
lattice. Our results are compared with mean-field like approaches. Based on
these former theories, we give a new formula to estimate the Curie temperature
of the system.

PACS number(s): 75.10H; 75.20E; 75.40M
\vskip 40pt

\vfill \eject

\section*{\S1. Introduction}

Binary Ising systems were studied from both bond and site perspectives
(Katsura and Matsubara 1974; Thorpe and McGurn 1978). In the bond-disordered
model the lattice sites are considered to be equivalent and the interaction
energies between the neighbouring sites are assigned randomly from a set of
possible values. In the site-disordered model the lattice sites are randomly
occupied by two different types of magnetic ions, $A$ and $B$, with spins
$S_A$ and $S_B$, the interaction parameters between neighbouring spins being
completly determined by their species. In this way there exist three type of
exchange interactions $J_{AA}$, $J_{AB}$ and $J_{BB}$, between  neighbouring
spins in the system. The disorder can be considered either quenched or
annealed.
The annealed systems are much more easier handled by mean-field like
approaches,
and so  there are better understood than the quenched ones. In spite of this,
for practical applications the quenched systems are much more appropriate. This
is the main reason why we proposed to limit the discussion  only for the case
of quenched systems.

In the case of only ferromagnetic interactions between the spins, binary
Ising or Heisenberg models were used with succes to describe the magnetic
properties of magnetic alloys (Vonsovski 1974; Luborsky 1980). When
antiferromagnetic and ferromagnetic interactions compete, frustration appears,
and the system becomes a Mattis-Luttinger type spin-glass model ( Binder and
Young 1986; Tatsumi 1977-78).

For real physical cases the site-disordered
models  are much more characteristic, and so we proposed to study the Ising
version of this model on the square lattice. We also considered the simplest
case of $S_A=S_B=\frac{1}{2}$, and all exchange interactions of ferromagnetic
type. The hamiltonian of the problem will be:
\begin{equation}
H=-\sum_{<i,j>}[J_{AA}\cdot \delta_{iA} \cdot \delta_{jA} + J_{BB} \cdot
\delta_{iB} \cdot \delta_{jB} + J_{AB} \cdot (\delta_{iA} \cdot
\delta_{jB} + \delta_{iB} \cdot \delta_{jA})]\cdot S_i^z \cdot S_j^z,
\end{equation}
where $\delta_{ix}=1$ if the spin $i$ is of type $x$, and $0$ otherwise.
The sum in (1) is refering to all nearest neighbours of the lattice. In
this paper we consider the two-dimensional version of the model. For results
concerning the three-dimensional, real model we have just preliminary
results (Neda 1994[b]).
The first molecular-field approximations for the proposed systems were due to
Vonsovskii (1940 and 1948). The frustrated systems were first studied by
Aharony (1975) using renormalization-group technics and by Tatsumi (1977-78)
with Monte Carlo simulations. The interesting case for us, with all the
interactions
of ferromagnetic type, was studied using a mean-field like approach by
Kouvel (1969), and with the coherent potential approximation by Foo and Wu
(1972). Mean-field theoretical approaches were also made in the works of
Tahir-Kheli and Kawasaki (1977) , respective Thorpe and McGurn (1978).
Ishikawa and Oguchi (1978) considered a Bethe-Peierls approach, and in the
work of Honmura, Khater, Fittipaldi and Kaneyoshi (1982) we find an
effective-field theory for the model. Monte Carlo studies for the critical
temperature of binary, ferromagnetic Ising alloys in function of the relative
species concentration and relative interaction energy between unlike ions
were performed by Scholten (1985) on the square lattice. Scholten
(1989) also studied the phase diagram for three-dimensional frustrated systems
on simple-cubic lattices, including next-nearest neighbour interactions too.
The phase diagrams of binary Ising ferromagnets were studied by Thorpe and
McGurn (1978), both in the site-disordered and bond-disordered cases. They
realized that these phase diagrams can be usefully cataloged in terms of the
initial slope $\frac{\partial \ln{T_c}}{\partial q}$ of the transition
temperature $T_c$ plotted in function of concentration $q$, at the two points
$q=0$ and $q=1$. Using the perturbation theory, they also determined the
initial slopes for two-dimensional systems. The phase diagrams for binary
Ising systems with randomly distributed exchange parameters were studied by
Kaneyoshi and Li (1987) using effective-field theory with correlations.
In the book from Vonsovskii (1974) and in the paper from Luborsky (1980), one
may find promising comparisions between experimental data and mean-field like
predictions. Diluted systems  ($J_{AB}=0$ and $J_{BB}=0$) also presented
interest for physicists (Wu 1982; Belokon and Semkin 1992; Neda 1994[a]).
Recently
there has been much interest in systems of mixed $S_A$ and $S_B$ spins, where
$S_A \neq S_B$ (Kaneyoshi 1989; Silva and Salinas 1991; Kaneyoshi,
Jascur and Tomczak 1992; Zhang and Yang 1993).

Although Monte Carlo
simulations were performed by Scholten (1985) on the considered model,
there remained some not completly clarified questions even in the simplest
two-dimensional and ferromagnetic case. The main problems are concerning the
values of the critical exponents and the elaboration of a practically usable
and general formula to estimate the critical temperature of the system.
Our work is intended to complete Scholtens paper in some sense, studiing
by a high-accuracy Monte Carlo simulation the Curie temperature of the
system. We do this in a review context, comparing our simulation results
with available theoretical formulas. In this manner we give a practically
usable and easy method of approximating the Curie temperature of the system
and illustrate the validity and limitations of different theoretical
approaches.

\section*{\S2. Used theoretical formulas}

The localized model of ferromgnetism involving nearest-neighbour exchange
integrals has an attractive simplicity for describing some magnetic systems.
Although this approach for the magnetism in metallic systems is not completly
acceptable due to the partially itinerant nature of the magnetic electrons,
the obtained results are usually in good agreement with experimental data. In
the case of binary magnetic alloys we are in a similar situation. The localized
model based on the Heisenberg or Ising hamiltonian (1) with nearest-neighbour
exchange, or the molecular-field theories proved to be applicable in
describing the variation of the critical temperature in function of the alloys
composition.

The first formula based on the molecular-field approximation
was derived, as we stated earlier, by Vonsovskii (1940; 1948) and used with
success to describe transition temperatures of binary magnetic alloys. The
proposed formula was:
\begin{equation}
T_c(q)=T_c(A,A)-2 \cdot [T_c(A,A)-T_c(A,B)]\cdot q + [T_c(A,A)+T_c(B,B)-
2 \cdot T_c(A,B)] \cdot q^2 ,
\end{equation}
where $T_c(A,A)$ and $T_c(B,B)$ are the Curie temperatures of the pure $A$ and
$B$ systems, $T_c(A,B)$ is the Curie temperature for a pure system caracterized
with all exchange interactions  equal with the ones between the $A$ and $B$
magnetic ions ($J_{AB}$), $T_c(q)$ is the Curie temperature of the mixture,
and $q$ is the concentration of the $B$ component.

We mention here that the critical temperature $T_c$ for an Ising system
on the square lattice, caracterized with $J$ exchange interaction
constants (considering just nearest-neighbour interactions) is given by
$T_c\approx 2.2681 \cdot \frac{J}{k_B}$, with $k_B$ the Boltzmann constant.

Using a phenomenological model based on mean-field theory suitably modified,
so that the individual atomic moments are allowed to vary in magnitude with
their local environment, and considering only nearest-neighbour interactions
Kouvel (1969) proposed the formula:
\begin{eqnarray}
 & T_c(q)=\frac{1}{2} \cdot [T_c(A,A) \cdot (1-q)+T_c(B,B) \cdot q] +
\nonumber \\
 & + \{ \frac{1}{4}\cdot
[T_c(A,A) \cdot  (1-q) - T_c(B,B) \cdot q]^2+ T_c(A,B)^2 \cdot q \cdot
(1-q) \} ^{\frac{1}{2}} .
\end{eqnarray}

In the work of Foo and Wu (1972) the disordered composition dependent
exchange interaction is treated in a coherent potential approximation (CPA).
In the limit of weak scattering their method give the mean-field like results,
but in the strong scattering limit they predict such effects as critical
concentration for the appearance of ferromagnetism in the diluted models
(Neda 1994[a]), which is not obtained in mean-field theories. They proposed the
following cubic equation for $T_c(q)$
\begin{eqnarray}
 & \alpha^2 \cdot T_c(q)^3 + \nonumber \\
 & +[\alpha \cdot (T_c(A,A)+T_c(B,B)+T_c(A,B))- \alpha
\cdot (1+\alpha) \cdot <T_c>] \cdot T_c(q)^2+ \nonumber \\
 & + [(1+\alpha) \cdot T_c(A,A) \cdot
T_c(B,B) \cdot T_c(A,B) \cdot < \frac{1}{T_c} > - \nonumber \\
 & -\alpha \cdot (T_c(A,A) \cdot
T_c(B,B) + T_c(A,B) \cdot T_c(A,A) + T_c(A,B) \cdot T_c(B,B))] \cdot T_c(q) -
\nonumber \\
 & -T_c(A,A) \cdot T_c(B,B) \cdot T_c(A,B)=0,
\end{eqnarray}
where
\begin{equation}
\alpha=\frac{z}{2}-1,
\end{equation}
with $z$ the coordination number of the lattice (in our case $z=4$), and
\begin{eqnarray}
 & <T_c>=(1-q)^2 \cdot T_c(A,A) + 2 \cdot q \cdot (1-q) \cdot T_c(A,B) +
q^2 \cdot T_c(B,B) ,\\
 & <\frac{1}{T_c}>= \frac{(1-q)^2}{T_c(A,A)}+ \frac{2\cdot q \cdot
(1-q)}{T_c(A,B)}
+ \frac{q^2}{T_c(B,B)}.
\end{eqnarray}

We mention that there are also other, more evoluate possibilities of
calculating
the Curie temperature, based on the Ising model (1) of the system, such as
mean-field like renormalization-group technics, series expansion and
perturbation methods. Unfortunately these are all very technical ones, and do
not give practically usable formulas.

\section* {\S3. The computer simulation method}

The Monte Carlo simulations performed by Scholten on the proposed model were
made by using the classical single spin-flip Metropolis algorithm (Metropolis,
Rosenbluth and Teller 1953). Due to this, his simulations were very
time-consuming , and he studied just a few choices for the values of the
interaction parameters ($\frac{J_{AB}}{J_{AA}}=0,1,2,4$ and $\frac{J_{BB}}
{J_{AA}}=4$). He also worked on relatively small $40\times 40$ square lattices
with periodic boundary conditions. For each choice of the interaction
parameters value he studied three cases for the concentration of the $B$
component $q=0.25$, $q=0.5$ and $q=0.75$ (the cases $q=0$ and $q=1$ are
evident). He compared his results with the ones obtained by (Tahir-Kheli
{\em et al.} 1977), (Thorpe {\em et al.} 1978), (Ishikawa {\em et al.}
1978) and (Honmura {\em et al.} 1982).

We proposed to continue Scholtens work by reconsidering the problem with
high-accuracy Monte Carlo simulations, using the more powerful cluster-flip
Swendsen and Wang method (Swensen, Wang and Ferrenburg 1992) with an original
recursion type algorithm. We considered many choices for the values of the
$J_{AB}$ and $J_{BB}$ interaction parameters, the value of $J_{AA}$ being
fixed. We proposed to compare our results obtained for the Curie temperature
with the ones given by equations (2), (3) and (4).

Our simulations  were performed on relatively large lattices, up to
$200 \times 200$ lattice sites. The critical temperature was obtained by
detecting the maximum in the fluctuation of the absolute value of
magnetization.
To achieve statistical equillibrium we considered up to $1000$ cluster-flips
and then studied the fluctuation for 2000 more iterations. The sensitivity for
the determination of the critical temperature was in general of the order
of $0.01\cdot T_c(A,A)$. For every chosen set of the interaction parameters
we covered the $q\in (0,1)$ concentration interval uniformly with $9$ to
$19$ simulation points. The program was written in C and the simulations
were performed on a CRAY Y-MP4D/464 computer and IBM R-6000 RISC workstations.

\section*{\S4. Results}

Our Monte Carlo results concerning the variation of the Curie temperature in
function of the B components concentration for the proposed two-dimensional
model are plotted with various symbols on figures 1 and 3-7. The curves
indicate theoretical results given from equations (2) and (4). In Fig. 1
considering four choices for the $J_{AB}$ interaction parameters ($J_{AA}$
and $J_{BB}$ fixed), we compare our Monte Carlo data with results given
by equation (2). In Fig. 2 we show some preliminary results for the
three-dimensional (simple-cubic) case, obtained with the same interaction
parameters as in Fig. 1, in comparision with the curve given by the (2)
molecular-field approximation. As one would expect it, we can also observ
that in the real, higher dimensional case the considered molecular-field
approximation is workin better.  From Fig. 1 we get, that on the square
lattice formula (2) predicts much higher results for the Curie temperature
than the real values. We checked that equation (3) predict even higher
values than (2). In Fig. 3  we show the same Monte Carlo data as in Fig. 1
in comparision with results obtained from equation (4). From Fig. 1 and 2
we conclude that in the considered cases the real critical temperatures are
limited by the two curves obtained from equations (2) and (4). In addition
to this, in Fig. 4 we show that almost a perfect fit with the realistic Curie
temperatures can be obtained, if we use the arithmetic mean of the
$T_c(q)$ values obtained from (2) and (4).

In Fig. 5-7 we tried to prove
our previous statements. So, we considered other choices for the
exchange interaction parameters, and thus for the $T_c(A,A)$, $T_c(B,B)$ and
$T_c(A,B)$ critical temperatures. We illustrated with thin dashed lines the
results obtained from equation (2) and (4) (dense dashes correspond to the
curve calculted from (4)). The continuous darker curve shows the arithmetic
mean of the $T_c(q)$ obtained from (2) and (4). We conclude again that in
general the values given from equations (2) and (4) limit nicely the
realistic simulation data, and their arithmetic mean gives a good estimate
for the Curie temperature. This arithmetic mean have stronger differnces with
our Monte Carlo data in the case when the $J_{AB}$ exchange interaction
parameters does not belong to the interval limited by the $J_{AA}$ and
$J_{BB}$ values.

\section*{\S5. Conclusions}

Our first conclusion is that the Curie temperatures calculated from
equations (2), (3) and (4) are not performant approximations of the real
values. However as expected, our simulations on simple-cubic lattices
reveal that the same (2) molecular-field approach is giving much better
results in the real three-dimensional case (Fig. 2).

For the case of the square lattice, generally the curves obtained for
the critical temperature from equations (2) and (4) limit rather nicely the
real values. Our Monte Carlo simulations indicate, that a few exceptions could
be for the small ($q\rightarrow 0$) and big ($q\rightarrow 1$)
concentration limit, when the
$J_{AB}$ interaction parameter is far from the interval limited by the values
of $J_{AA}$ and $J_{BB}$.

Our most important conclusion is that, the theoretical curve constructed as
the arithmetic mean of the Curie temperatures obtained from equations (2) and
(4) proved to be a good approximation for the critical temperature of a
binary Ising ferromagnet on the square lattice.

Similar preliminary results for the three-dimensional case are given
in a recent preprint (Neda 1994[b]).

\section*{Acknowledgements}

\samepage

This study was finished during a bursary offered by the Norwegian Research
Council. We thank Y. Brechet, A. Coniglio,
L. Csernai, and L. Peliti for their continuous help and useful discussions.

\newpage

\section*{References}

{\bf A}harony, A.H., 1975, {\em Phys. Rev. Lett.}, {\bf 34}, 590 \\
{\bf B}elokon, V.I., and  Semkin, S.V., 1992, {\em Sov. Phys. JETP} {\bf 75},
	       680 \\
{\bf B}inder, K., and Young, A.P., 1986, {\em Rev. Mod. Phys.} {\bf 58}, 801 \\
{\bf F}oo, E-Ni, and Wu, Der-Hsueh, 1972, {\em Phys. Rev. B} {\bf 5}, 98 \\
{\bf H}onmura, R., Khater, A.F., Fittipaldi, I.P., and Kaneyoshi, T., 1982,
	      {\em Solid State Commun.} {\bf 41}, 385  \\
{\bf I}shikawa, T., and  Oguchi, T., 1978, {\em J. Phys. Soc. Jpn.} {\bf 44},
	       1097 \\
{\bf K}aneyoshi, T., and  Li, Z.Y., 1982, {\em Phys. Rev. B} {\bf 35}, 1869 \\
{\bf K}aneyoshi, T., 1989, {\em Phys. Rev. B} {\bf 39}, 12134 \\
{\bf K}aneyoshi, T., Jascur, M., and  Tomczak, P., 1992, {\em J. Phys. Condens.
		Matter.} {\bf 4}, L653 \\
{\bf K}atsura, S., and  Matsubara, F., 1974, {\em Can. J. Phys.} {\bf 52},
	      120 \\
{\bf K}ouvel, J.S., 1969, in {\em  Magnetism and Metallurgy vol. II.}, eds.
	      A.E. Berkowitz and E. Kneller (Academic Press) \\
{\bf L}uborsky, F.E., 1980, {\em J. Appl. Phys.} {\bf 51}, 2808 \\
{\bf M}etropolis, N., Rosenbluth, A.W., Rosenbluth, M.N., Teller, A.H.,
	       and Teller, E., 1953, {\em J. Chem. Phys.} {\bf 21}, 1087 \\
{\bf N}\'eda, Z., 1994, {\em J. Phys. I. (Paris)} (to be published in
	       february issue ) \\
{\bf N}\'eda, Z., 1994, {\em Scientific Report } {\bf 02/1994}
(Univ. of Bergen, Norway) \\
{\bf S}cholten, P.D., 1985, {\em Phys. Rev. B} {\bf 32}, 345 \\
{\bf S}cholten, P.D., 1989, {\em Phys. Rev. B} {\bf 40}, 4981  \\
{\bf S}ilva, N.R. da, and  Salinas, S.R., 1991, {\em Phys. Rev. B} {\bf 44},
	       852 \\
{\bf S}wendsen, R.H.,  Wang, J.S., and  Ferrenburg, A.M., 1992, in {\em  The
	       Monte-Carlo Method in Condensed Matter Physics}, ed. K. Binder
			     (Springer-Verlag) \\
{\bf T}ahir-Kheli, A., and  Kawasaki, T., 1977, {\em J. Phys. C} {\bf 10},
	      2207 \\
{\bf T}atsumi, T., 1977,  {\em Prog. Theor. Phys.} {\bf 57}, 1799 \\
{\bf T}atsumi, T., 1978,  {\em Prog. Theor. Phys.} {\bf 59}, 1428;
	      {\bf 59} 1437 \\
{\bf T}horpe, M.F., and McGurn, A.R., 1978, {\em Phys. Rev. B} {\bf 20},
	       2142
\newpage
$\:$ \\
{\bf V}onsovskii, S.V., 1940, {\em Dokl. Akad. Nauk. SSSR} {\bf 26}, 364 \\
{\bf V}onsovskii, S.V., 1948, {\em Zhurn. Tekh. Fiz.} {\bf 18}, 131 \\
{\bf V}onsovskii, S.V., 1974,  {\em Magnetism II.}, (John Willey)
	      pp. 776-822 \\
{\bf W}u, F.Y., 1982, {\em Rev. Mod. Phys.} {\bf 54}, 235 \\
{\bf Z}hang, G.M., and  Yang, C.Z., 1993, {\em Z. Phys. B} {\bf 91}, 145 \\

\newpage
\section*{Figure Captions}
\vspace{0.10in}

$\:$ \\
\vspace{0.15in}

{\bf Fig. 1} Monte Carlo results for the variation of the Curie temperature
as a function of the $B$ components concentration for four choices of the
$T_c(A,B)$ critical temperature. Solid curve is given by equation (2).
\vspace{.25in}

{\bf Fig. 2} The same plot as in Fig. 1 for simulations done on the
simple-cubic lattice.
\vspace{.25in}

{\bf Fig. 3} The Monte Carlo results from Fig.1 in comparision with the
Curie temperatures obtained from (4).
\vspace{.25in}

{\bf Fig. 4} The Monte Carlo results from Fig. 1 in comparision with the
arithmetic mean of the Curie temperatures obtained from (2) and (4).
\vspace{.25in}

{\bf Fig. 5} The dots and triangles represents Monte Carlo simulations for the
given $T_c(A,B)$ critical temperatures. The thin dashed lines indicate the
results obtained from formulas (2) and (4) (dense dashes correspond to (4)).
The dark continuous line indicate the arithmetic mean obtained from (2) and
(4).
\vspace{.25in}

{\bf Fig. 6} The case when we have no exchange interactions between the atoms
of the $B$ component ($J_{BB}=0$) and $T_c(A,A)=T_c(A,B)=100$. Dots are
Monte-Carlo results and the curves have the same meaning as in Fig. 4.
\vspace{.25in}

{\bf Fig. 7} Monte Carlo results (dots) for $T_c(A,A)=T_c(B,B)=100$  and \\
$T_c(A,B)=500$. The curves represents the same as in Fig. 4.

\end{document}